\documentclass[twocolumn]{aastex631}
\usepackage{amsmath}

\submitjournal{ApJ}

\shorttitle{$\gamma$-ray emission of the SNR 29.37+0.1}
\shortauthors{Zheng et al.}

\newcommand{\snr}{G29.37+0.1}
\newcommand{\hess}{HESS J1844$-$030}
\newcommand{\gr}{$\gamma$-ray}
\newcommand{\grs}{$\gamma$-rays}
\newcommand{\fermi}{{\it Fermi}}

\newcommand{\gamb}{\gamma_{\rm b}}

\newcommand{\kep}{K_{\rm ep}}

\begin{document}

\title{4FGL~J1844.4$-$0306: high-energy emission likely from the supernova remnant G29.37+0.1}

\correspondingauthor{Zhongxiang Wang, Xiao Zhang}
\email{wangzx20@ynu.edu.cn, xiaozhang@nju.edu.cn}

\author{Dong Zheng}
\affiliation{Department of Astronomy, School of Physics and Astronomy, Yunnan
University, Kunming 650091, China}

\author[0000-0003-1984-3852]{Zhongxiang Wang}
\affiliation{Department of Astronomy, School of Physics and Astronomy, Yunnan
University, Kunming 650091, China}
\affiliation{Shanghai Astronomical Observatory, Chinese Academy of Sciences, 80 Nandan Road, Shanghai 200030, China}

\author{Xiao Zhang}
\affiliation{School of Astronomy \& Space Science, Nanjing University,
163 Xinlin Avenue, Nanjing 210023, China}
\affiliation{Key Laboratory of Modern Astronomy and Astrophysics, Nanjing University, Ministry of Education, China}

\author{Yang Chen}
\affiliation{School of Astronomy \& Space Science, Nanjing University,
163 Xinlin Avenue, Nanjing 210023, China}
\affiliation{Key Laboratory of Modern Astronomy and Astrophysics, Nanjing University, Ministry of Education, China}

\author{Yi Xing}
\affiliation{Shanghai Astronomical Observatory, Chinese Academy of Sciences, 80 Nandan Road, Shanghai 200030, China}



\begin{abstract}
	Very-high-energy (VHE) observations have revealed approximately 100 
	TeV sources in our Galaxy, and a significant fraction of them are
	under investigation for understanding their origin. We report
	our study of one of them, HESS~J1844$-$030. It is found possibly 
	associated with the supernova remnant (SNR) candidate G29.37+0.1, and
	detailed studies of the source region at radio and X-ray frequencies 
	have suggested that this SNR is a composite one, containing a pulsar 
	wind nebula (PWN) powered by a
candidate young pulsar. As the GeV source 4FGL~J1844.4$-$0306 is also located
in the region with high positional coincidence, we analyze its $\gamma$-ray 
	data obtained with the Large Area 
	Telescope on-board the {\it Fermi Gamma-ray Space Telescope }. 
	We determine the GeV $\gamma$-ray emission is extended,
	described with a Log-Parabola function.
The obtained spectrum can be connected to that 
of the VHE source HESS J1844$-$030. Given the properties and those from
multi-frequency studies, we discuss the origin of the $\gamma$-ray emission
	by considering that the two \gr\ sources are associated.
Our modeling indicates that while the TeV part would have either a hadronic
	(from the SNR) or a leptonic origin (from the putative PWN),
	the GeV part would arise from a hadronic process. Thus we conclude
	that 4FGL~J1844.4$-$0306 is the likely GeV counterpart to G29.37+0.1.
\end{abstract}



\section{Introduction} \label{sec:intro}

Supernova remnants (SNRs) and their homologues pulsar wind nebulae (PWNe)
are conspicuous sources in our Galaxy, as both types of the sources can be 
bright from frequencies radio to \grs, showing features of different
physical processes (e.g., \citealt{rey17,gs06}). 
In SNRs, their shockfronts are believed to accelerate both 
protons and electrons, and thus SNRs are considered as 
the sites from which Galactic cosmic rays (CRs) originate 
(e.g., \citealt{bz34,byk+18}). If there is is dense material for 
the accelerated protons to run into, considerable high-energy and 
very-high-energy (VHE)
\gr\ emissions are generated through proton-proton collisions, pion production
and subsequent pion decay (e.g., \citealt{der86,dav94}), i.e., 
the so-called hadronic process. The accelerated electrons can
also generate emissions through synchrotron 
and inverse Compton Scattering (ICS) processes (e.g., \citealt{stu+97}), namely 
the so-called leptonic process. A PWN is powered by its central pulsar that
drives an ultrarelativistic magnetized wind. The wind, consisting of 
electron/positron pairs (leptons), interacts with the ambient medium
and slows down at the so-called termination shock, where the ram pressure 
of the wind is balanced by the internal pressure of the PWN 
(see \citealt{gs06}, for a review).
It is believed that leptons within the PWN are accelerated to 
relativistic energies
via diffuse shock acceleration or other mechanisms still being investigated
at the termination shock. 
The accelerated particles then escape into the nebula where they radiate 
from radio to VHE gamma-rays via synchrotron and ICS processes 
\citep{kc84,aa96,zcf08,gsz09,tt11}. 
Therefore the SNRs and PWNe
are targets of high-energy studies for understanding mechanisms of
particle acceleration and related radiation, and are expected to be associated
with VHE TeV sources found in our Galaxy.

In recent years, many TeV sources located in the Galactic 
plane have been detected in VHE surveys \citep{hess18,3hawc}, and presumably
most of them should be SNRs or/and PWNe. Indeed for example, among 78 sources 
detected in the High Energy Spectroscopy System (H.E.S.S.) Galactic plane
survey, 16 and 12 were
identified as SNRs and PWNe respectively \citep{hess18}.
However on the other hand, a significant fraction
of these TeV sources do not have obvious counterparts at other wavelengths, 
suggesting more
detailed studies should be carried out for clarifying their possible SNR or
PWN origin.

The TeV source \hess\ is one of them \citep{hess18}, while its position
coincides with that of a SNR candidate G29.37+0.1. This candidate was first
revealed from 
the radio imaging survey of the first Galactic quadrant \citep{hel+06}, and
appears to have an interesting S-shaped radio structure in the center,
surrounded by a diffuse halo. Because of the 
positional coincidence, 
multi-wavelength follow-up studies of 
the source region were conducted 
by \citet{cas+17}. They found that the S-shaped structure is likely
a background radio galaxy and the weak radio halo could be the 
shell of a composite SNR (Figure~\ref{fig:ts}). They also suggested that
\hess\ could be associated with the radio galaxy. However,
\citet{pet19} re-analyzed
the archival data and proposed that the TeV source could be the 
VHE counterpart to an X-ray nebula source G29.4+0.1, which is likely 
a PWN powered by a potential pulsar 
(X-ray source PS1)
(see details in \citealt{cas+17,pet19}; see also Figure~\ref{fig:ts}).

As part of our studies for understanding the origins of several tens of 
the Galactic TeV sources that do not have obvious counterparts at other
wavelengths
\citep{hess18,3hawc}, we have noted that there is a \gr\ source in the GeV 
band detected with the Large Area Telescope (LAT; \citealt{atw+09})
on-board {\it the Fermi Gamma-ray Space Telescope (Fermi)}. 
This source appeared in the \fermi\ LAT first source catalog 
(1FGL; \citealt{1fgl}) as 1FGL~J1844.3$-$0309c and in the fourth catalog (4FGL;
\citealt{4fgldr3}) as 4FGL~J1844.4$-$0306. The source has a position
consistent with those of \hess\ and G29.37+0.1 (Figure~\ref{fig:ts}). We 
thus analyzed the
\fermi\ LAT data. The results provide a more complete picture for the \snr\ 
region. Here we report the analysis and results. In Section~\ref{sec:ana},
we describe our analysis and present the corresponding results. In 
Section~\ref{sec:dis}, we discuss the implication of the results by 
modeling the broadband spectral energy distribution (SED) that involves
the sources observed at multi-wavelengths in the region of \snr.

\section{\fermi\ LAT Data Analysis and Results}
\label{sec:ana}
\subsection{LAT Data and Source Model}
\label{sec:LAT data}

We used the \fermi-LAT Pass 8 data from 2008-08-04 15:43:36 (UTC) to 2021-12-01 
00:00:00 (UTC). The region of interest (RoI) was set with a size of 
$15^{\circ} \times 15^{\circ}$ centered at the position of 4FGL J1844.4$-$0306. 
The events were selected with parameters evclass=128 and evtype=3 and the
maximum apparent zenith angle at 90 degree. Good time interval were calculated 
with the tool {\tt gtmktime} by setting the filter conditions 
DATA\_QUAL$>$0 \&\& LAT\_CONFIG=1 for Galactic point source analysis, 
and data in good time intervals were selected to be used in our 
analysis.

We built a source model using the data release 3 of 4FGL (4FGL-DR3), 
which was
based on the analysis of 12-year LAT data (\citealt{4fgldr3}). 
All sources within 15 degree radius of 4FGL J1844.4$-$0306 were included in
the source model. For the sources within (outside) a 5-degree distance
from the RoI center,  
their the spectral parameters were allowed to vary (were frozen at the 
4FGL-DR3 values). In addition, the Galactic background 
and extragalactic diffuse emission models, 
gll\_iem\_v07.fits and iso\_P8R3\_SOURCE\_V3\_v1.txt respectively,
were included in the source model. The normalizations of these two 
components were always set as free parameters in the following 
analyses.
\begin{table}
	\begin{center}
\caption{Likelihood analysis results with the LP, PL, and PLSEC models}
	\label{tab:model}
	\begin{tabular}{lcccc}
	\hline
	Model & Best-fit parameters & $\log(L)$ & TS \\ 
	\hline
		LP$^\dagger$ & $\alpha = 2.91 \pm 0.12$ & &\\
		& $\beta = 0.30 \pm 0.08$ & & \\
	LP$^\ddagger$  & $\alpha = 2.59 \pm 0.04$ & 10744073.48 & 406 \\
		& $\beta = 0.26 \pm 0.03$ & & \\
	PL$^\ddagger$  & $\Gamma = 2.66 \pm 0.03$  & 10744065.59 & 456 \\
	PLSEC$^\ddagger$ & $\Gamma_{S} = 2.34 \pm 0.04$ & 10744072.31 & 416\\
		&	$d=0.29\pm0.04$ &  & \\
	LP$^\ast$ & $\alpha = 2.40\pm0.03$ & 10744092.98 & 595 \\
		& $\beta = 0.04\pm0.02$ & & \\
	PL$^\ast$ & $\Gamma = 2.66 \pm 0.03$ & 10744092.43 & 613\\
\hline
\end{tabular}
\end{center}
	\footnotesize{$^\dagger$Values given in 4FGL-DR3; $^\ddagger$Values for a point source; $^\ast$ Values for an extended source with radius 0\fdg3.}
\end{table}

\subsection{Likelihood analysis} 
\label{subsec:lh}

\begin{figure*}
   \centering
   \includegraphics[width=0.33\textwidth, angle=0]{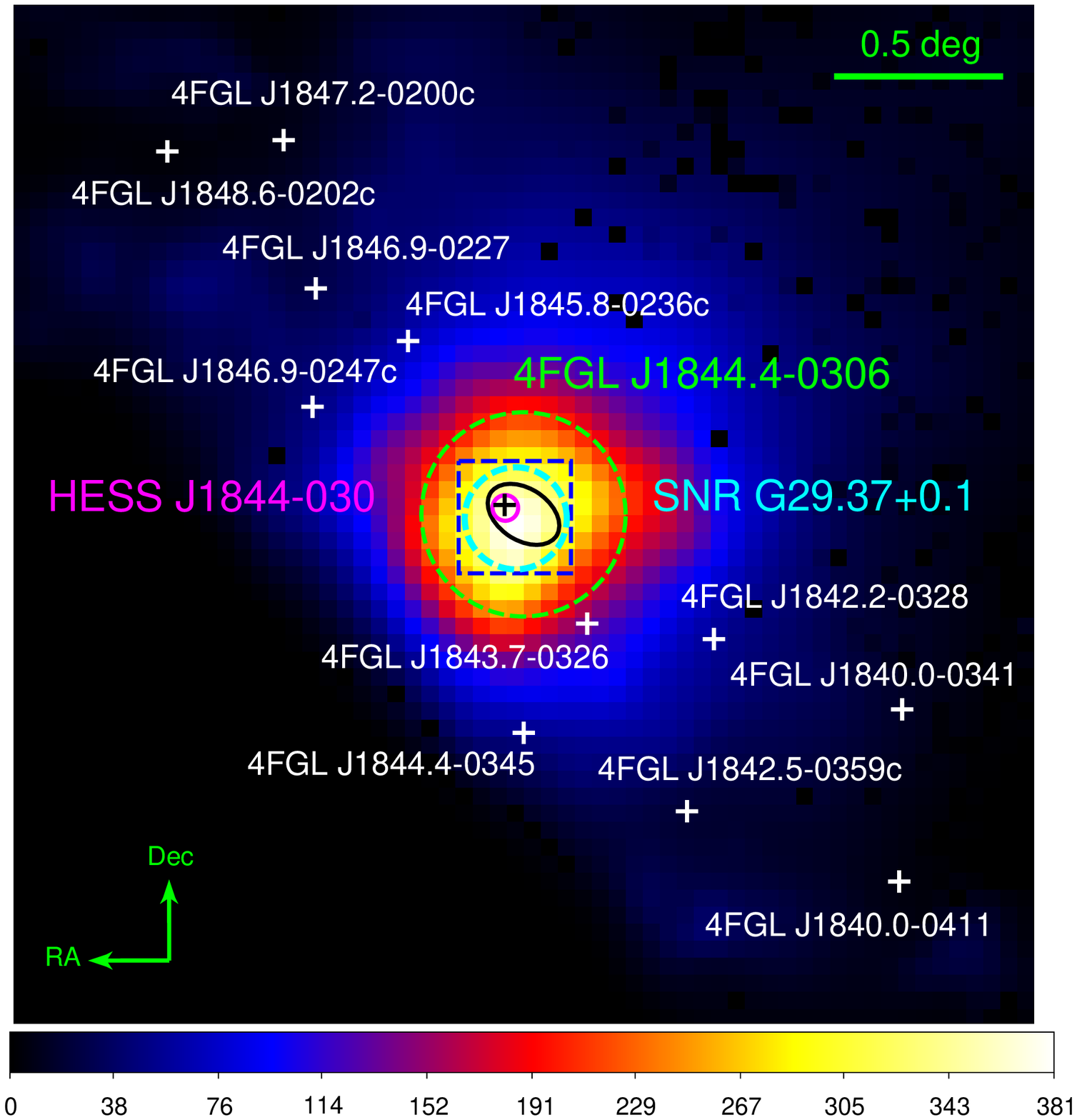}
   \includegraphics[width=0.336\textwidth, angle=0]{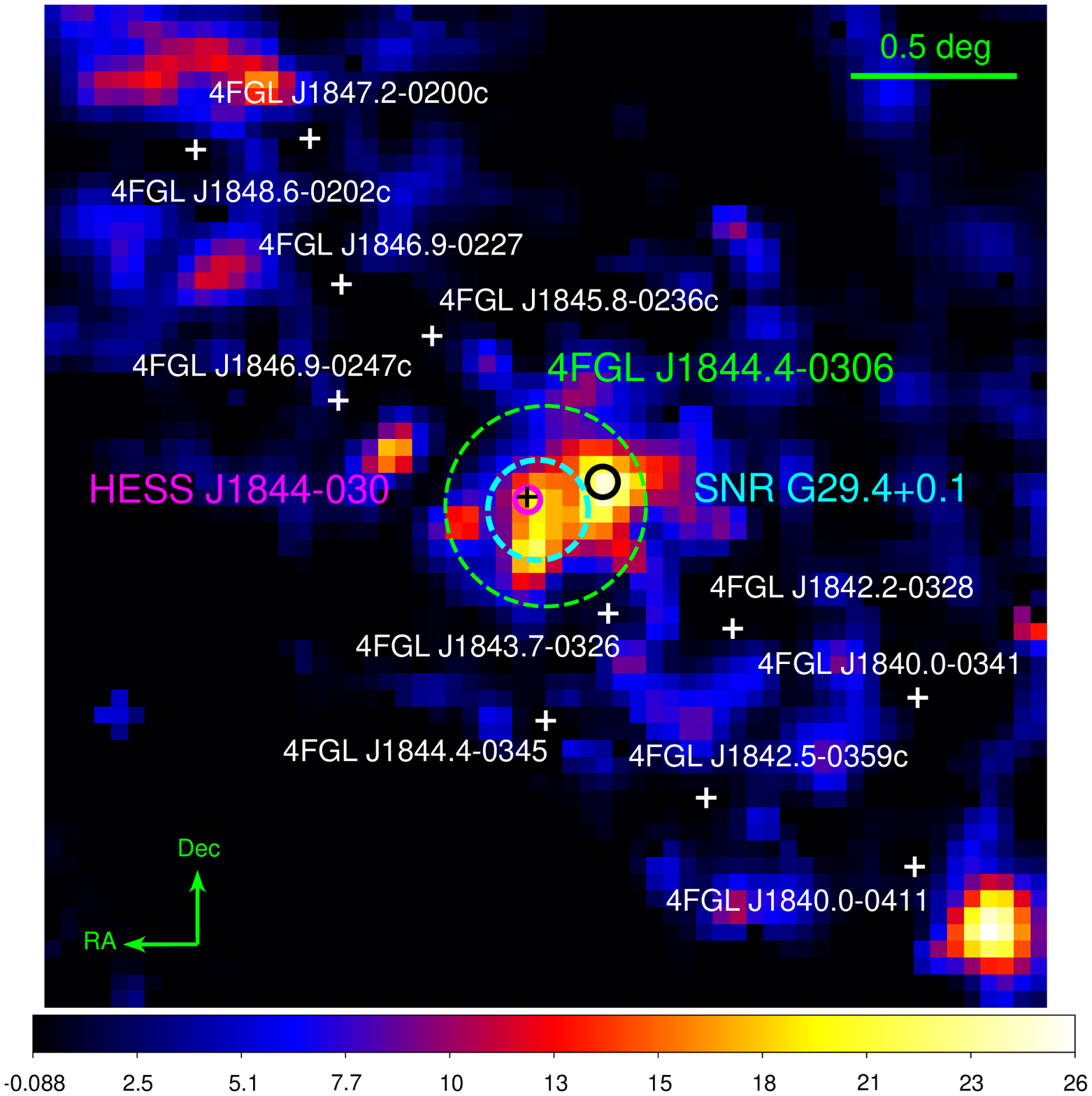}
   \includegraphics[width=0.33\textwidth, angle=0]{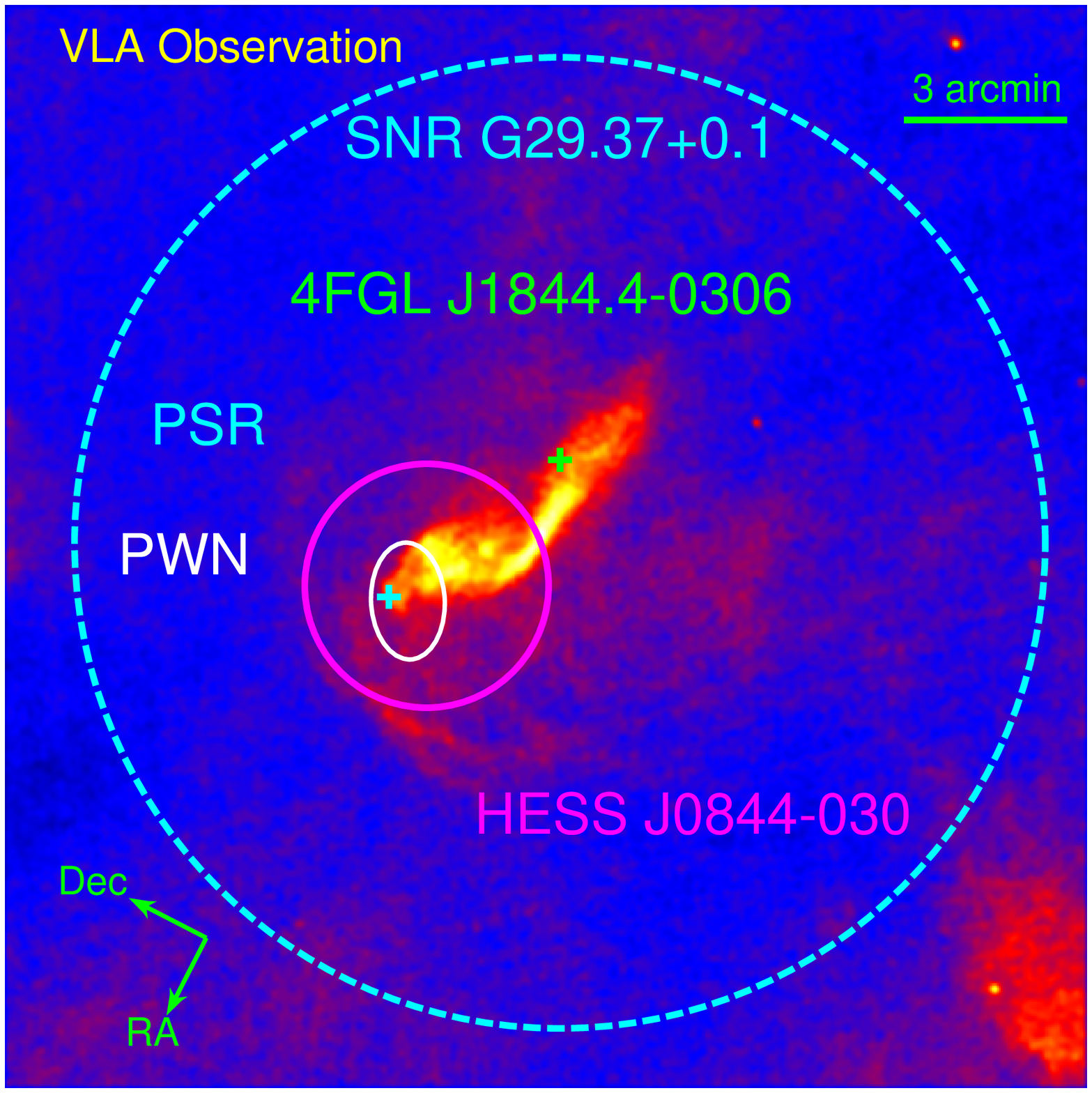}
	\caption{{\it Top:} TS maps of the $3^{\circ}\times 3^{\circ}$ 
	region centered at the target 4FGL~J1844.4$-$0306 in 0.3--500\,GeV 
	({\it left}) and 5--500\,GeV ({\it right}). 
All the catalog sources except the target are removed in the map.
	The magenta circle is the 95\% error 
	region of HESS J1844$-$030 (see also the {\it bottom} panel),
the cyan dashed circle is the radio region of G29.37+0.1, and
	the green dashed circle marks the 0\fdg3 radius extension 
	(Section~\ref{subsec:ext}).  In addition, the position of 
	a candidate pulsar \citep{cas+17} is marked as a black plus. 
	In the {\it left}, the black ellipse is the 95\% error region 
	of the target, and in the {\it right}, the black circle marks 
	the position determined for the $\geq 5$\,GeV additional residual 
	emission (Section~\ref{subsec:lh}).
	{\it Bottom:} 
	Very-Large-Array radio image of G29.37+0.1 \citep{hel+06}, whose
	size is indicated by the blue dashed square in the {\it top left} 
	panel. 
	The S-shaped structure is clearly visible in the central region, 
	surrounded by weak diffuse emission (i.e., the halo) that 
	is also visible at 610\,MHz \citep{cas+17}. 
	The candidate pulsar and PWN \citep{cas+17,pet19} 
	are marked by a cyan plus and a white ellipse respectively. 
	The position of 4FGL~J1844.4$-$0306 (marked by a green plus) is 
	nearly at the center of G29.37+0.1, and is 0\fdg062 in R.A. and
	0\fdg028 in Decl. away from the pulsar.  
	\label{fig:ts}}
   \end{figure*}

The source 4FGL J1844.4$-$0306 has an unknown type in 4FGL-DR3
as a point source (PS), having a position of R.~A. = 281\fdg119,
Decl. = $-$3\fdg116 (J2000.0) with an error ellipse (at a 95\% confidence)
semi-major and semi-minor axes of 0\fdg116 and 0\fdg076 respectively
(Figure~\ref{fig:ts}).  The source's emission is fitted
with a Log-Parabola (LP) model in the catalog, 
$dN/dE = N_{0}(E/E_{b})^{-[\alpha + \beta\log(E/E_{b})]}$, where 
the $\alpha$ and $\beta$ catalog values are given in 
Table~\ref{tab:model} ($E_b=1.6$\,GeV is fixed). 
To fully investigate the \gr\ properties of this source,
we also considered a power-law (PL) model, $dN/dE = N_{0}(E/E_{0})^{-\Gamma}$,
and a sub-exponentially cutoff power law (PLSEC; \citealt{4fgldr3}),
$dN/dE = N_{0}(\frac{E}{E_{0}})^{\Gamma_{S}-\frac{d}{2}\ln\frac{E}{E_{0}}-\frac{db}{6}\ln^{2}\frac{E}{E_{0}}-\frac{db^{2}}{24}\ln^{3}\frac{E}{E_{0}}}$ 
(when $\vert b\ln\frac{E}{E_{0}} \vert < 10^{-2}$; see \citealt{4fgldr3} for
details). For the latter, we intended
to test if the emission could be better described with that of a pulsar,
and the function form was recently proposed to be used for pulsars 
so as
to reduce the correlation between the parameters in analysis \citep{4fgldr3}.

A standard binned likelihood analysis was performed to data in 0.3--500 GeV,
in which the image scale was 0\fdg1 and the size in pixels was
150.  We did not use the data below 0.3\,GeV because the source is in
the Galactic plane where the background emission is dominant below the energy
\citep{4fgl20}. The source position was fixed at that given in
4FGL-DR3, and we first assumed a PS spatial model as in
the catalog. Each of the three spectral models described above was 
tested in the analysis. 
In the LP model, $E_b$ is the scale parameter and was always fixed at 1.6\,GeV
in our analysis, following that given given in 4FGL-DR3.
The similar scale parameter $E_{0}$ in the PL model was 
fixed at 1\,GeV. 
The parameters $b$ and $E_{0}$ in the PLSEC model were fixed
at $2/3$ and 1\,GeV respectively, where the latter was set because 
of the correlation 
between $E_0$ and the normalization.
The resulting best-fit parameters and 
TS values are given in Table~\ref{tab:model}.
Of the PS spectral tests, the PL provided the largest TS 
value.
However when we compared the likelihood values from the analyses with 
the models,
using $\Delta$TS$\simeq \sqrt{-2 log(L_{i}/L_{j})}$ where $L_{i/j}$ are
the maximum likelihood 
values from model $i$ and $j$ (see Table~\ref{tab:model}),
the obtained $\Delta$TS values indicated that the LP
and PLSEC models were  $\sim$4.0$\sigma$ and $\sim$3.7$\sigma$, respectively,
more preferable than
the PL model. Below we considered the LP model as the one for describing
emission from J1844.4$-$0306. It can be noted that the parameter 
$\alpha$ of the LP model we obtained is smaller than that given in the catalog,
but within the $3\sigma$ uncertainty (Table~\ref{tab:model}).


To illustrate the source field in the GeV band, we calculated a TS map
in 0.3--500\,GeV. By removing all the sources except J1844.4$-$0306 
in the region, the TS map was obtained and is shown in the top 
left panel of Figure~\ref{fig:ts}.
The source is in a crowded region and relatively bright among the nearby 
sources. As can be seen, its 95\% positional error region is highly coincident 
with that (95\% confidence) of HESS J1844$-$030 (magenta circle) and the source 
region of G29.37+0.1 (cyan dashed circle), where the latter was
given by the radio observation \citep{pet19}.
We also show a VLA (1.4\,GHz) radio image of G29.37+0.1 
\citep{hel+06} in the bottom panel of Figure~\ref{fig:ts}. 
The position of 
J1844.4$-$0306 is close to the SNR's center, at which the bright S-shaped 
structure is also located. The \gr\ position is approximately 0\fdg068 away
from the X-ray source PS1 (as well as the putative PWN).
The offset is within the 95\% confidence error region 
of the \gr\ position, and thus the \gr\ source is in possible association
with either PS1/PWN or the SNR.

In addition, given the 0.3--500\,GeV TS map
shows possible evidence for source extension, we further calculated TS maps
in the energy ranges of 5--500, 8--500, and 10--500\,GeV to check. In the
top right panel of Figure~\ref{fig:ts}, we show the 5--500\,GeV one. Additional
TS$\sim$20 emission northwest of J1844.4$-$0306 is seen, and it turns to be
the only visible residual emission in 10--500\,GeV.
We ran \textit{gtfindsrc} to the 10--500\,GeV data, and a position 
of R.A.=280\fdg948, Decl.=$-$3\fdg044 (equinox J2000.0), with a 1$\sigma$ 
nominal uncertainty of 0\fdg046, was obtained for the residual emission. 
However based on our analysis conducted in
the next section, the emission could not be identified
as an individual source.
   
\subsection{Spatial analysis}
\label{subsec:ext}

\begin{figure}
   \centering
   \includegraphics[width=0.46\textwidth, angle=0]{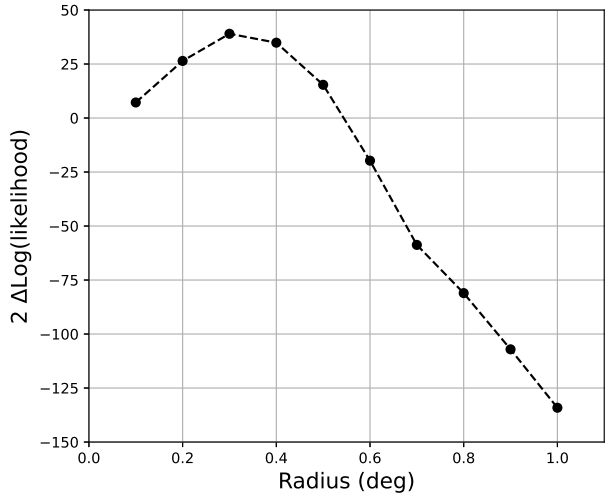} 
   \caption{Results from comparing the likelihood values of considering
	an extended source with that of a PS for 4FGL J1844.4$-$0306. 
At radius $0\fdg3$, the source is found to be most significantly extended.
	\label{fig:ext}}
   \end{figure}

\begin{figure}
   \centering
   \includegraphics[width=0.46\textwidth, angle=0]{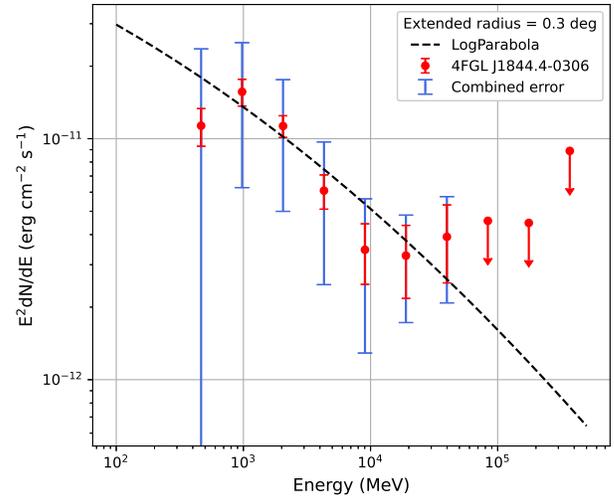} 
   \caption{\gr\ spectrum in 0.3--500\,GeV for 
	an extended source (with radius $0\fdg3$) at the position of
	of 4FGL J1844.4$-$0306, for which the systematic uncertainties
	(Table~\ref{tab:spec}) provided
	in \citet{eag22} were incorporated. The best-fit LP model is shown 
	as a dashed line.
	\label{fig:espec}}
   \end{figure}

Given the above analysis results, we checked
whether or not J1844.4$-$0306 is extended by setting up a uniform disk model 
in the source model file.
A range of $0\fdg1$--$1\fdg0$ radius, with a step of $0\fdg1$, for the uniform 
disk was tested.
Assuming a LP spectral model for the source and
performing the likelihood analysis, 
each of the resulting likelihood values $L_{\rm ext}$ was compared with that 
$L_{\rm ps}$ from considering a PS, that is, $2\log(L_{\rm ext}/L_{\rm ps})$.
The results are shown in Figure~\ref{fig:ext}. At radius $0\fdg3$, the 
difference of the log-likelihood values is the largest, 38.4. The value indicates that
the source is extended at a significance of 6.2$\sigma$ 
(given by $\sqrt{2\log(L_{\rm ext}/L_{\rm ps})}$).
Using the $0\fdg3$ uniform disk model for the source, the best-fit LP 
parameters were found to be $\alpha = 2.40\pm0.03$, $\beta = 0.04\pm0.02$, 
and the 0.3--500\,GeV photon flux 
$=3.08\pm0.15$\,photon\,cm$^{-2}$\,s$^{-1}$ (with a TS value of 595; 
also
given in Table~\ref{tab:model}). 
The flux is higher than that from considering a PS by $\sim 40$\%.
Also it can be noted that since the $\beta$ 
value is close to zero, the model fit has small curvature and is nearly a 
power law (Figure~\ref{fig:espec}). We tested a PL spectral model to
fit the extended emission
and obtained $\Gamma=2.66\pm0.03$, with a very similar $\log L_{\rm ext}$ value
but a larger TS value (see Table~\ref{tab:model}). Thus a PL model equally
well describes the extended emission.

Because the TS map at $\geq$5\,GeV shows additional emission
northwest of J1844.4$-$0306 (Figure~\ref{fig:ts}), we conducted tests to
determine whether the higher energy emission could be considered as
an individual source. Adding it in the source model as a PS with
its position at the one obtained in Section~\ref{subsec:lh}, we
performed the likelihood analysis by assuming a PS or a uniform disk 
(with the same radius setup as the above)
for J1844.4$-$0306 (i.e., two PSs or an extended disk plus a PS in the
source model).
Likely due to the faintness of the higher energy emission (TS$\sim$20), 
the resulting likelihood values were not significantly increased. We thus
concluded that the additional emission could not be determined as another
source and it may be considered as part of the extended emission of
J1844.4$-$0306.

We extracted a \gr\ spectrum of J1844.4$-$0306 in 0.3--500\,GeV by considering
an extended source with radius $0\fdg3$.
The energy range was divided into 10 evenly logarithmically spaced energy
bins. The maximum likelihood analysis was performed to the data in each
energy bin, in which the sources within 5 degree of the target were set to
have a free normalization parameter and all the other parameters of the sources
in the source file were fixed at the values obtained in the above likelihood
analysis assuming the LP spectral model for our source. 
The obtained spectral data points are given in 
Table~\ref{tab:spec} and shown in Figure~\ref{fig:espec},
for which the fluxes with TS$\geq$4 were taken as 
measurements and otherwise 95\% flux upper limits were derived. 

We note that \citet{eag22} also conducted a study of 
4FGL~J1844.4$-$0306, and similar results to ours were obtained, although 
the details were slightly different. For example, the best-fit template
in the study was a radial Gaussian function, which assumes a centrally
peaked intensity profile, rather than a flat-disk one in our analysis.
Nevertheless, the source extension was
found to be $\sim0\fdg3$ from the standard deviation of the Gaussian function,
the same as ours.
The spectrum of the source and the statistical uncertainties were reasonably 
consistent with ours.
In addition, the systematic uncertainties were obtained in the study, and 
importantly, likely due to the crowdedness of the field, the values are
mostly larger than the statistical ones. We thus obtained the systematic
uncertainties for our spectral data points by interpolating from their values
(as the energy bins in the study were 7 over an energy range of 0.3--2000\,GeV; 
\citealt{eag22}). The obtained values (Table~\ref{tab:spec}) were added in 
quadrature to the statistical uncertainties, and the combined uncertainties
are
shown in Figure~\ref{fig:espec}, for which it can be noted that the systematic
uncertainties are dominant at low, $<10$\,GeV energies and the first one 
at 0.5\,GeV is large, comparable to the flux value (Table~\ref{tab:spec}). 
The spectrum and the statistical plus systematic 
uncertainties are taken as the final measurements for the source
(and to be discussed in Section~\ref{sec:dis}),
but the first
data point is removed due to the large systematic uncertainty.

Based on these analyses, we summarize that the source J1844.4$-$0306,
at the position R.~A. = 281\fdg119, Decl. = $-$3\fdg116 (J2000.0) with an 
error ellipse (at a 95\% confidence) semi-major and semi-minor axes of 
0\fdg116 and 0\fdg076 respectively, has \gr\ emission being extended with 
a circular
radius of 0\fdg3 and described with a LP (or a PL) spectral model. Weak 
$\geq 5$\,GeV emission was seen northwest of it, but the additional emission 
could not be significantly detected as another source.

\subsection{Timing and variability analysis}
\label{subsec:tva}

As the X-ray source PS1 has been suggested as a putative young 
pulsar \citep{cas+17,pet19}, we performed timing 
analysis to the LAT data of J1844.4$-$0306 to search for possible $\gamma$-ray 
pulsations. The LAT events within an aperture radius of 0.5\,deg in 
0.1--500\,GeV were selected for the analysis. We divided the whole
data into sets of 2-yr long, and the timing-differencing blind search 
technique \citep{atw+06} was applied to the events in each 2-yr data 
set. The search range of frequency ($\nu$) and frequency derivative over 
frequency ($\dot{\nu}/\nu$) were 0.5--32\,Hz and
0--1.3$\times$10$^{-11}$\,s$^{-1}$ (where the high-end range value is 
that of the Crab pulsar),
with steps of 1.90735$\times$$10^{-6}$\,Hz and 
9.451$\times$$10^{-16}$\,s$^{-1}$,
respectively. However, no significant $\gamma$-ray pulsations from the source 
was found.
\begin{table}
	\caption{Spectral data points for 4FGL~J1844.4$-$0306}
\begin{tabular}{lcr}
\hline
$E$   & $E^{2}dN(E)/dE$ & TS \\
(GeV) & ($10^{-11}$ erg cm$^{-2}$ s$^{-1})$ & \\
\hline
0.5    & $1.13 \pm 0.20\pm1.21$ & 71 \\
1.0    & $1.56 \pm 0.20\pm0.92$ & 207\\
2.1    & $1.13 \pm 0.12\pm0.62$ & 157\\
4.3    & $0.61 \pm 0.10\pm0.35$ & 50\\
9.0    & $0.35 \pm 0.10\pm0.19$ & 15\\
19.0   & $0.33 \pm 0.11\pm0.11$ & 11\\
39.9   & $0.39 \pm 0.14\pm0.12$ & 11\\
83.7   & $\geq 0.46$ & 0.2 \\
175.8  & $\geq 0.45$ & 0.1 \\
369.1  & $\geq 0.89$ & 1.4 \\
\hline
	\label{tab:spec}
\end{tabular}

	\footnotesize{Statistical and systematic uncertainties for fluxes are
	given, where the latter are interpolated from those given 
	in \citet{eag22}.}
\end{table}

Since the S-shaped structure is likely an extragalactic radio galaxy
and such sources could emit variable $\gamma$-rays \citep{4fgldr3}, we 
also checked 
the variability of J1844.4$-$0306. In 4FGL-DR3, its 
{\tt Variability\_Index} is given to be 8.93, lower than the variability
threshold value 24.725 (for 12 bins with 1\,yr per bin; \citealt{4fgldr3}). 
We re-calculated {\tt Variability\_Index} by using the best 
extended model we found above, and our data contain
13 1-yr time bins. A light curve consisting of 13 bins
was obtained, in which only the spectral 
normalizations of all sources within 5 degree of J1844.4$-$0306 
were set as free parameters. The obtained {\tt Variability\_Index} was 9.30.
Thus no variations were detected in the source, and there is no strong
evidence to suggest an association between the candidate radio galaxy and 
the \gr\ source.


\section{Discussion and Summary}
\label{sec:dis}

Because of the positional coincidence of 4FGL~J1844.4$-$0306 
with \snr\ (as well as \hess), we analyzed the \fermi\ LAT data for 
this source. When considering it as a PS, its 
emission appears to be described with a curved function, as more preferably 
fitted with an LP or PLSEC model than a PL. 
The emission could arise from the putative pulsar (i.e., PS1 in 
\citealt{cas+17,pet19}), whose X-ray emission has been studied in detail
by \citet{pet19}. At a distance of $\sim$6.5\,kpc \citep{pet19}, 
the \gr\ luminosity would be $\sim 2.5\times 10^{35}$\,erg\,s$^{-1}$, suggesting
a \gr\ efficiency of $\sim$2.6\% based on the spin-down energy $L_{\rm sd}$
estimated for 
the pulsar by \citet{pet19}. The efficiency value appears to be in the proper
range derived
for young pulsars (characteristic ages less than $\sim10^6$ yrs;
\citealt{abd+13,lwx21}). However aside from these, no clear evidence,
such as pulsed emission from the putative pulsar 
(e.g., Section~\ref{subsec:tva}), has been found.

Instead, we have found that the \gr\ source
is extended, at a significance level of 6.2$\sigma$. Also the model 
fit for the extended
source was changed to have small curvature ($\beta\simeq 0.04$; 
Table~\ref{tab:model}), different from those of pulsars that often have an 
exponential cutoff at several GeV energies \citep{abd+13}.
In addition, its GeV band spectrum can be 
relatively well connected to that of \hess\ in TeV energies (see below), likely
indicating its association with \hess. 
Given these properties and those derived from multi-wavelength data analyses, 
we discuss its possible origin in the following sub-sections by considering 
an SNR scenario (Section~\ref{subsec:snr}), 
a PWN scenario (Section~\ref{subsec:pwn}), or a 
composite PWN-SNR scenario (Section~\ref{subsec:com}). A summary is 
provided at the end in Section~\ref{subsec:sum}.

\begin{deluxetable*}{ccccccccc}[htb]
\tablecaption{Summary of parameters used in the SNR scenario. \label{tab:para_snr}}
\tablecolumns{9}
\tablewidth{0pt}
\tablehead{
  \colhead{Model} &
  \colhead{$K_{\rm ep}^a$} &
  \colhead{$n_{\rm t}^a$} &
  \colhead{$\alpha$} &
  \colhead{$W_{\rm e}^b$} &
  \colhead{$W_{\rm p}$} &
  \colhead{$E_{\rm c,e}$} &
  \colhead{$E_{\rm c,p}^a$} &
  \colhead{$B^a$} \\
  \colhead{ } &
  \colhead{ } &
  \colhead{ (${\rm cm}^{-3}$) } &
  \colhead{ } &
	  \colhead{ ($10^{47}$ erg) } &
  \colhead{ ($10^{48}$ erg) } &
  \colhead{ (TeV) } &
  \colhead{ (PeV) } &
  \colhead{ ($\mu$G) }
}
\startdata
 A & 0.01 & 10   & 2.6 & 24.0 & 240.2 & 40 & 1 & 5\\
 B & 0.1  & 10   & 2.6 & 74.1 &  74.7 & 10 & 1 & 5 \\
 C & 0.01 & 1000 & 2.5 & 0.2 & 2.1 & 50 & 1 & 5\\
 D & 0.1  & 1000 & 2.5 & 0.8 & 0.8 & 50 & 1 & 5\\
\enddata
\tablenotetext{a}{Fixed in the fitting process.}
\tablenotetext{b}{Calculated based on $K_{\rm ep}$, $\alpha$, and $W_{\rm p}$.}
\end{deluxetable*}

\subsection{SNR Scenario}
\label{subsec:snr}

Taking the radio halo of \snr\ as a Galactic SNR \citep{cas+17,pet19}, 
the high-energy and VHE \gr\ emissions from it could correspond
to 4FGL~J1844.4$-$0306 and \hess\ respectively. Note that there are no radio
flux measurements and no X-ray detection of the halo, the latter possibly
due to high hydrogen column density ($\sim 10^{23}$\,cm$^{-2}$) towards
the target region \citep{cas+17,pet19}.
We thus model the GeV fluxes obtained in this work and TeV ones from 
H.E.S.S. \citep{hess18}. 
Both hadronic and leptonic processes are considered,
in the latter of which the non-thermal bremsstrahlung process is included. 

We assume that the particles
accelerated by the SNR shock have a power-law form with a high-energy cutoff,
\begin{equation}
dN_i/dE_i = A_i (E_i/ 1\ {\rm GeV})^{-\alpha_i}{\rm exp}(-E_i/E_{\mathrm{c},i})\ \ \ ,
\end{equation}
where $i=$~e or p, $\alpha_i$ is the power-law index, $E_{\mathrm{c}, i}$ is 
the cutoff energy, and the normalization $A_i$ is determined by the total 
energy in particles with energy above 1\,GeV, $W_i$.
To model the broadband SED from these 
energy distributions of particles, we used 
the PYTHON package Naima \citep{naima}, which includes the 
synchrotron \citep{akp10}, non-thermal bremsstrahlung \citep{bergg99}, 
ICS \citep{kak14}, and pion-dacay \citep{katv14} processes.
In the calculation, we set $\alpha=\alpha_{\rm e}=\alpha_{\rm p}$ presuming
the charge-independent acceleration process.  Considering the long lifetime 
of protons and the lack of the constraint on the high-energy cutoff, 
$E_{\rm c,p}=1$\,PeV is set. For the electrons, the cooling-limited maximum 
energy is used if the cutoff energy can not be constrained by the data.
In addition, we employ the parameter $\kep=A_{\rm e}/A_{\rm p}$ instead of 
$W_{\rm e}$ to control the number ratio of electrons to protons at 1\,GeV.
The measurement of the local CRs around the Earth implies $\kep\sim 0.01$.
Some theoretical predictions suggest that $\kep$ may be up to  
$\sim$0.1 \citep{mbed17}.
Taking these into account, two cases, $\kep=0.01$ and 0.1, are respectively
considered.
\begin{figure*}
   \centering
   \includegraphics[width=0.9\textwidth, angle=0]{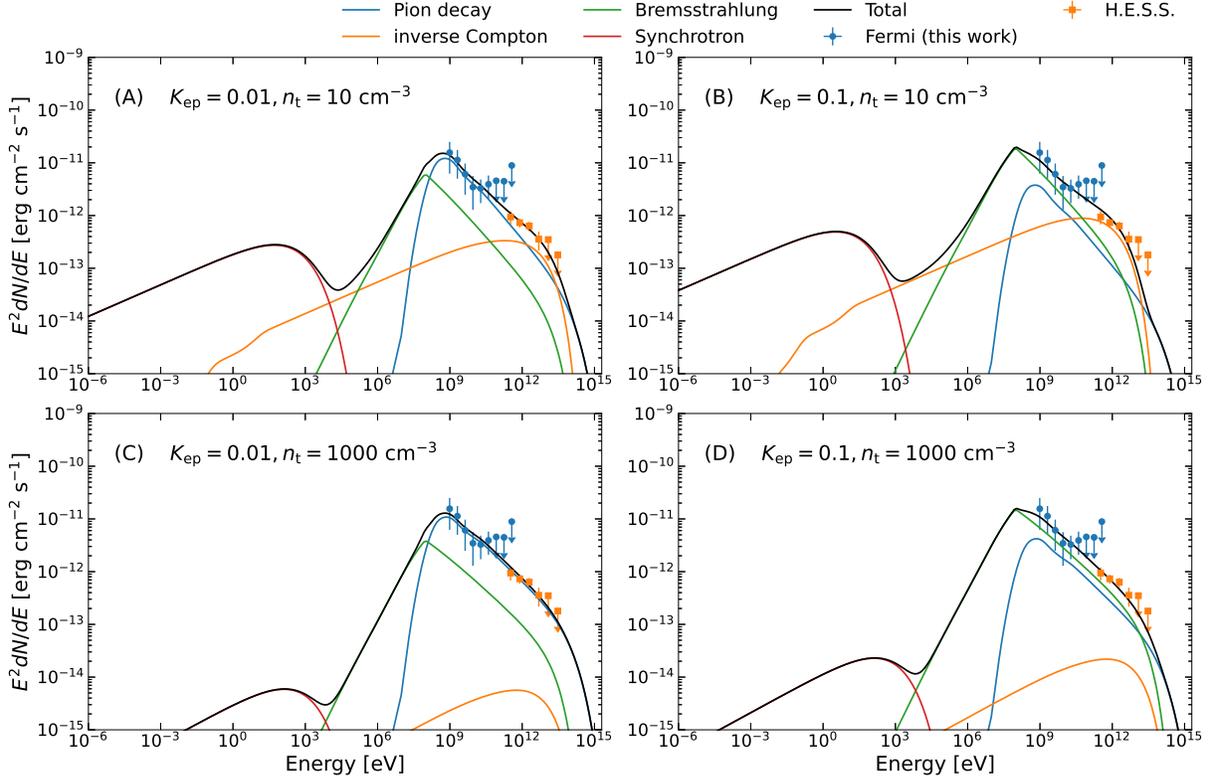} \\
	\caption{  Model fits in the SNR scenario (Section~\ref{subsec:snr})
	to the high-energy and VHE \gr\ spectrum, combined from that of
	4FGL~J1844.4$-$0306 with that of \hess, where the TeV data points are 
	from \citet{hess18}. 
	The model-fit parameters are listed in 
	Table~\ref{tab:para_snr}.
    \label{fig:sed_snr}}
\end{figure*}

For the number density of the target gas ($n_{\rm t}$), no reliable value can 
be derived according to the current observations of the target field.
\citet{pet19} gave a rough estimate of the ambient density 
$n_0\sim8\ {\rm cm}^{-3}$ based on the assumption that the SNR shock has swept 
part of the HI gas surrounding it.
While there has been no evidence supporting the interaction of the SNR 
with any molecular cloud (MC; \citealt{pet19}),
three MCs with a density $n_{\rm MC}$ (in atomic hydrogen) of order
$\sim10^3\ {\rm cm^{-3}}$ and a distance of 5--6\,kpc have
been found at the boundary of the radio halo \citep{cas+17}, which possibly
provide sufficient targets in the hadronic process.
Thus, we also consider two cases: $n_{\rm t}=10$ and $1000\ {\rm cm^{-3}}$.

The seed photon fields for the ICS process include the cosmic microwave 
background (CMB) radiation, the dust infrared (IR) with a temperature 
of 40\,K and an energy density of $1\ {\rm eV\ cm^{-3}}$, and the star
light (SL) with a temperature of 4000\,K and an energy density of 
$2\ {\rm eV\ cm^{-3}}$. The IR and SL components around the target's position 
are calculated based on the approximate method provided by \citet{sis11}.

To summarize our model setup, there are four cases named Model A--D 
(Table~\ref{tab:para_snr}), and in each case there are three free parameters: 
$\alpha$, $W_{\rm p}$, and $E_{\rm c,e}$.
The SED of each model that best fits the GeV-TeV fluxes is obtained 
(Figure~\ref{fig:sed_snr}) and the 
corresponding parameters are summarized in Table~\ref{tab:para_snr}.
For this SNR scenario, since there are no radio or X-ray flux measurements, 
we just present 
the synchrotron fluxes under the magnetic field strength of $B=5\ \mu$G.
In the case of $n_{\rm t}=10\ {\rm cm}^{-3}$, $E_{\rm c,e}$ can
be roughly constrained by the TeV fluxes, but for 
$n_{\rm t}=10^3\ {\rm cm}^{-3}$, it can not be and is thus set to
be the synchrotron cooling-limited maximum energy 
$\sim 50 (t_{\rm snr}/10^4\ {\rm yr})(B/5\ \mu{\rm G})^{-2}$~TeV, 
where $t_{\rm snr}\sim10^4$~yr, the SNR's age approximately estimated by
\citet{pet19}.

As can be seen in Figure~\ref{fig:sed_snr}, the radiation mechanism of 
the GeV-TeV \gr\ emission depends on the parameters $\kep$ and $n_{\rm t}$.
For example, the GeV-TeV emission has a leptonic origin for $\kep=0.1$ and 
$n_{\rm t}=10\ {\rm cm}^{-3}$ (Figure~\ref{fig:sed_snr}b).
While for $\kep=0.01$ and $n_{\rm t}=1000\ {\rm cm}^{-3}$ 
(Figure~\ref{fig:sed_snr}c), it has a pure hadronic origin.
Considering the fact that $\kep$ of the order of $\sim10^{-2}$ is commonly 
favored based on the measured electron-to-proton ratio in the CRs 
(e.g., \citealt{bla13}),
Model A and C are preferred. The modeling results suggest that
the GeV \gr\ emission may mainly has a hadronic origin,
and depending on the target-gas density, the TeV emission 
may be dominated by the hadronic process (when the density is high) or 
contains substantial contribution from the ICS process (when the density is
relatively low).

Considering the canonical explosion energy $10^{51}$\,erg
and typical energy conversion efficiency of 10\%, the product
$n_{\rm t}W_{\rm p}\sim 2\times 10^{51}$\,erg\,cm$^{-3}$ in Model A and C 
implies that \snr\ must be
in a high-density environment. Indeed, we note that the additional
$\geq$5\,GeV emission (cf., top right panel of 
Figure~\ref{fig:ts}) is outside of the SNR region but within 
the 0\fdg3-radius extension and is 
positionally coincident with the three MCs mentioned in \citet{cas+17}. 
We may suspect that at least part of the higher energy emission of 
J1844.4$-$0306 likely arises
from the interaction of the SNR or its precusor with the MCs, for the latter
of which theoretical studies have predicted \citep{fed+15}. Hopefully,
the SNR hadronic scenario, the
association between the SNR and the three MCs 
or other dense matter, could be clarified in future observational studies.
In addition, flux measurements on the radio emission can help to 
constrain the magnetic field, since it affects the synchrotron cooling-limited 
maximum energy and thus the ICS contribution to the TeV \gr\ emission.


\subsection{PWN scenario}
\label{subsec:pwn}

\begin{figure*}
   \centering
   \includegraphics[width=0.45\textwidth, angle=0]{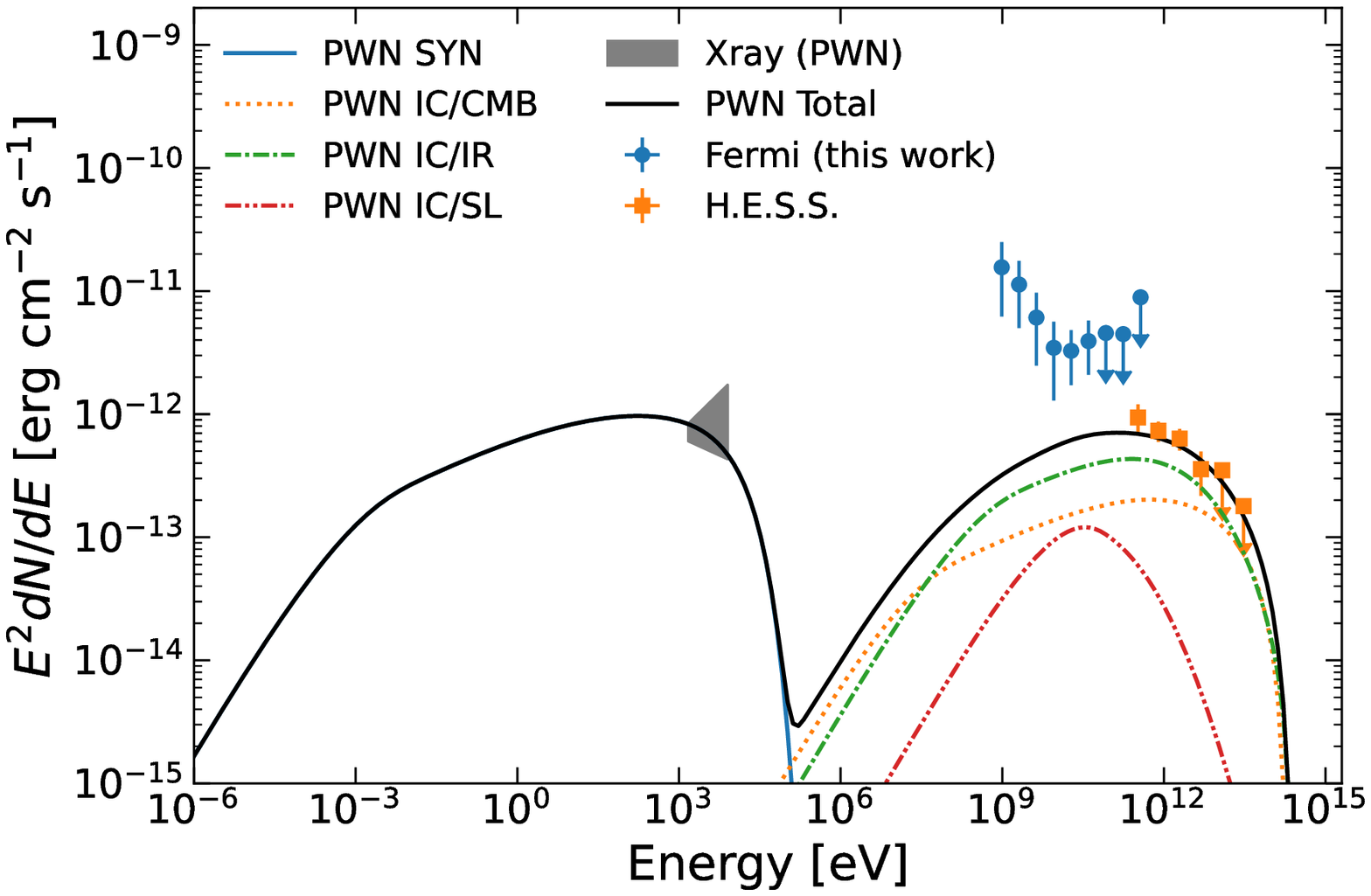}
   \includegraphics[width=0.45\textwidth, angle=0]{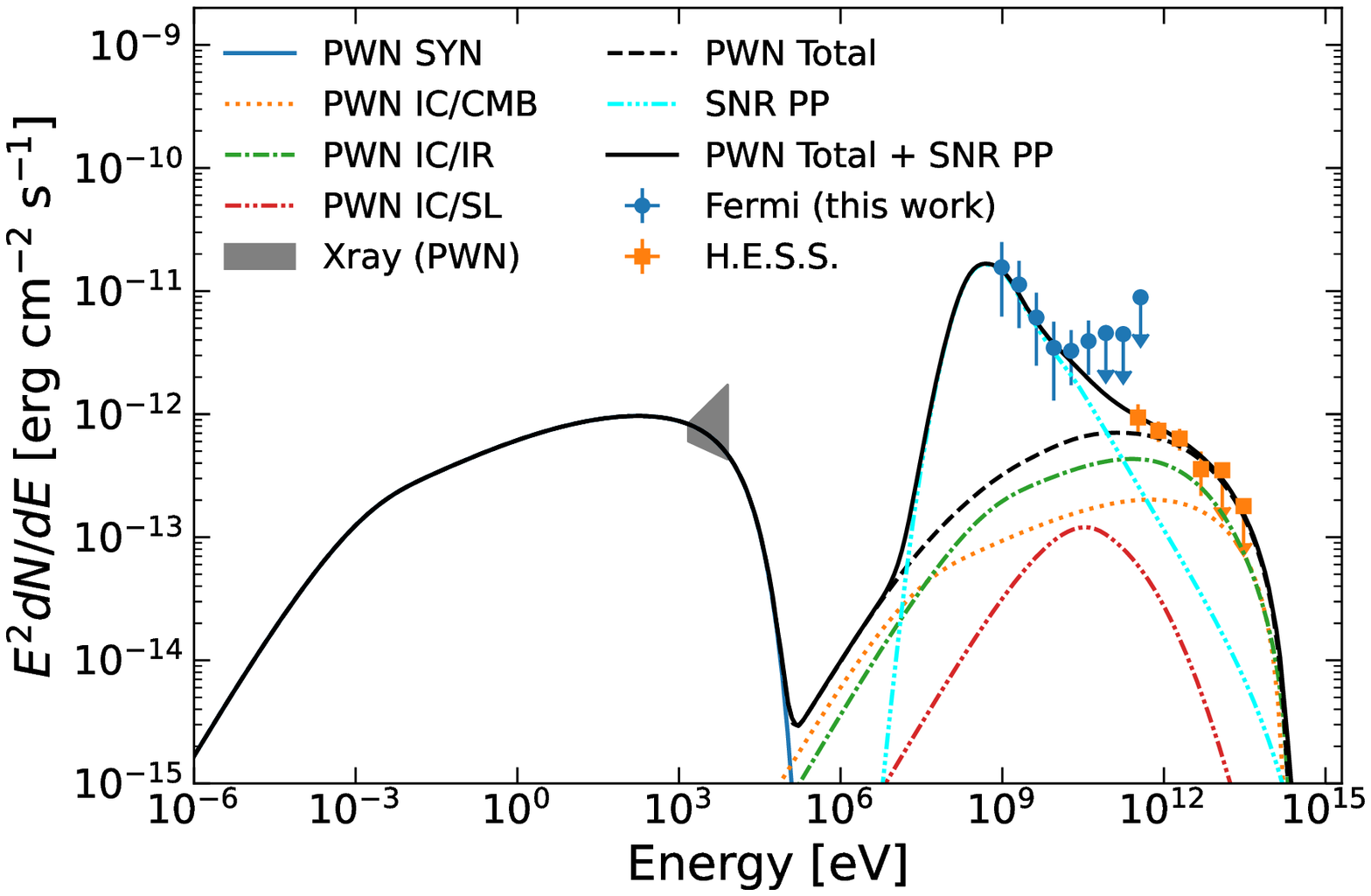}
	\caption{Same broad-band SED as that in 
	Figure~\ref{fig:sed_snr}, but with the X-ray model spectrum 
	in 1.5--8.0\,keV (obtained for the candidate PWN; \citealt{pet19}) 
	also shown (the black shaded region). 
	{\it Left} panel: model fit to the SED 
	in the PWN scenario 
	(Section~\ref{subsec:pwn}), which
	can not explain the GeV fluxes. {\it Right} panel: model fit to
	the SED in the composite PWN-SNR scenario (Section~\ref{subsec:com}). 
	Leptonic components from the putative PWN can fit the X-ray and
	TeV \gr\ fluxes, while the hadronic component (dash-dotted cyan
	line) from the SNR is needed to explain the GeV \gr\ fluxes.
    \label{fig:sed}}
\end{figure*}

Both \citet{cas+17} and \citet{pet19} have shown that the
non-thermal X-ray nebula surrounding PS1 likely has
a PWN origin, powered by PS1 the putative young pulsar. We thus also
investigate whether a PWN scenario can provide an explanation
for the \gr\ emissions. Here we consider a broad-band SED
that additionally includes
the model spectrum in 1.5--8.0\,keV for the X-ray nebula, which was
derived from multiple {\it Chandra} and {\it XMM-Newton} 
observations \citep{pet19}.

Based on the time-dependent PWN model \citep[e.g.,][]{zcf08,lcz10,tt10,mtr12},
the lepton spectrum, $N(\gamma,t)$, inside the PWN is governed by the 
continuity equation in the energy space
\begin{equation}
\label{eq:dis_evol}
\frac{\partial N(\gamma,t)}{\partial t} = -\frac{\partial}{\partial\gamma}[\dot{
\gamma}(\gamma,t)N(\gamma,t)] -\frac{N(\gamma,t)}{\tau(\gamma,t)} + Q(\gamma,t)\ ,
\end{equation}
where $\gamma$ is the Lorentz factor of the leptons, $\dot{\gamma}(\gamma,t)$
the energy loss rate including the synchrotron radiation, the ICS process, and the adiabatic loss \citep[see][]{lcz10},
and $\tau(\gamma,t)$ the escape time which can be estimated via Bohm
diffusion \citep[e.g.][]{zcf08}.
The injection rate of leptons $Q(\gamma,t)$ is generally assumed to be a
broken power-law function
\begin{equation}
\label{eq:Qinj}
Q(\gamma,t)=Q_0(t)\left\{
\begin{array}{ll}
 (\gamma/\gamb)^{-\alpha_1}\quad {\rm for}\ \gamma_{\rm min}<\gamma<\gamb \\
 (\gamma/\gamb)^{-\alpha_2}\quad {\rm for}\ \gamb<\gamma<\gamma_{\rm max}\ ,
\end{array}
\right.
\end{equation}
where $\gamb$ denotes the break energy, $\alpha_1$ and $\alpha_2$ are the
spectral indices below and above the break energy respectively.
The normalization is determined from the total energy in the leptons,
$(1-\eta)L_{\rm sd}$, converted from the spin-down power $L_{\rm sd}$
of the pulsar, where $\eta$ is
the energy conversion fraction from $L_{\rm sd}$ to that of the magnetic field.
The evolution of the spin-down power is governed by 
$L_{\rm sd}=L_0[1+(t/\tau_0)]^{-(n+1)/(n-1)}$, where $L_0$, $\tau_0$, 
and $n$ are the initial spin-down luminosity,  the initial spin-down timescale,
and the breaking index, respectively \citep{ps73}.  
The minimum energy of the injected leptons is assumed to be
$\gamma_{\rm min}=1$, while the maximum energy is constrained by introducing
a parameter $\varepsilon$, which requires that the Larmor radius of the leptons
must be less than the termination shock radius \citep[see][]{mtr12}. 
The maximum energy is thus given by
\begin{equation}
\gamma_{\rm max} = \frac{\varepsilon e \kappa}{m_e c^2}\sqrt{\eta\frac{L_{\rm sd}}{c}}\ \ \ ,
\end{equation}
where $e$ is the electron charge, $m_e$ is the electron mass, $c$ is speed of 
light, and $\kappa\approx3$ is the magnetic compression 
ratio \citep[][and reference therein]{mtr12}.  

The lepton energy distribution in the PWN can be obtained by solving
equation~\ref{eq:dis_evol}.
Then the spectrum from radio to X-ray frequencies is calculated via
the synchrotron process. The same population of the leptons also produce
the \gr\ emission from the
sub-GeV to VHE regime through the ICS process.
The seed photon fields for the ICS process are the same as those used in 
the SNR scenario.
The distance and age of the radio halo of \snr\ was constrained to be
$\sim$6.5\,kpc and $\ge10$~kyr respectively \citep{pet19}.
We assume that the putative pulsar, originating from the supernova
explosion, has the same distance and age ($t_{\rm psr}$).
For the pulsar's breaking index, we adopt the standard value of $n=3$
as due to the magnetic dipole radiation. The initial spin-down timescale
$\tau_0$ is
degenerate with the initial spin-down luminosity $L_0$ for
given $t_{\rm psr}$ and $n$. We assume $\tau_0=2$~kyr.
To explain the \gr\ luminosity,
$L_0=1.0\times10^{38}\ {\rm erg\ s^{-1}}$, corresponding to
$L_{\rm sd}=2.8\times10^{36}\ {\rm erg\ s^{-1}}$ at the present time.
The $L_{\rm sd}$ value is
roughly consistent with the range of $10^{36}-10^{37}\ {\rm erg\ s^{-1}}$
estimated from the empirical relations between the spin-down energy and the
X-ray emission given in \citet{pet19}.
Due to lack of the radio data, the index below the break $\gamma_b$ is 
fixed at a common value $\alpha_1=1.5$ in the calculation \citep{ss11}.
The other parameters are approximated to be
$\alpha_2=2.55$, $\gamb=2.5\times10^{5}$, $\eta=0.09$, and $\varepsilon=0.1$,
which are all consistent with those of typical \gr-bright PWNe \citep[see Table 1-3 in][]{zzf18}.
These parameters and used values are summarized in Table~\ref{tab:par}, and the corresponding SED is displayed in the left panel of Figure~\ref{fig:sed}.
At the present time, the average magnetic field in the PWN given 
by $\eta=0.09$ is 5.2\,$\mu$G, and the maximum energy of electrons constrained 
by $\varepsilon\approx 0.1$ is 250\,TeV.
As shown in the left panel of Figure~\ref{fig:sed}, however, the ICS process 
of the PWN can not provide a fit to the GeV \gr\ data points, suggesting that 
the pure PWN scenario can not simultaneously explain the observed
GeV and TeV \gr\ emissions. 


\subsection{Composite PWN-SNR scenario}
\label{subsec:com}

In the pure PWN model, the GeV data is not explained and needs an 
additional component that may be contributed by the SNR. We thus consider
a composite scenario, in which the role of the PWN is similar to that of
the ICS process in the SNR scenario. For simplicity, only the hadronic 
process discussed in Section~\ref{subsec:snr} is included. In addition,
to stress the role of the PWN,
we assume the TeV emission is dominated by the contribution of the PWN.  

To explain the GeV \gr\ emission, the
power-law index $\alpha_{\rm p}=2.75$ and the total energy in protons
$W_{\rm p}=3.5\times 10^{48}/(n_{\rm t}/10^3\,{\rm cm^{-3}})$\,erg are needed.
The resulting lepto-hadronic SED (i.e, PWN+SNR) is shown as the black solid
line in the right panel of Figure~\ref{fig:sed}. This hybrid model can thus explain the broad-band
SED, although it also requires a high-density environment
for providing targets in the hadronic process.

\subsection{Summary}
\label{subsec:sum}

Following the radio and X-ray studies of the region that contains the VHE
source HESS~J1844$-$030 and the SNR \snr\ conducted by \citet{cas+17} 
and \citet{pet19}, we have analyzed the \fermi\ LAT data for the GeV source
4FGL~J1844.4$-$0306 that is also located in the region.
The GeV source is determined to be extended, but whether
it contains emission from a young pulsar, whose existence has been 
suggested from the X-ray studies, remains to be resolved.
The GeV spectrum can be connected to that of 
HESS~J1844$-$030, suggesting that the \gr\ emission arises from
the composite SNR. By modeling the \gr\ spectrum and the broad-band
SED that include the X-ray flux measurements for the putative PWN, we show 
that while either a hadronic or a leptonic origin 
may explain the TeV emission, a hadronic process is likely required in order to
explain the GeV emission. Thus the \fermi\ LAT source 4FGL~J1844.4$-$0306
is likely the GeV counterpart to the SNR \snr, and our study helps provide
a clearer picture for understanding this complex Galactic region. 

\begin{deluxetable*}{llll}[htb]
\tablecaption{Summary of parameters used in the modeling of the PWN. \label{tab:par}}
\tablecolumns{4}
\tablewidth{0pt}
\tablehead{
  \colhead{Parameters} &
  \colhead{Symbol} &
  \colhead{Value} &
  \colhead{Ref.}
}
\startdata
Distance (kpc) & $d$ & 6.5 & \citet{pet19} \\
Age (kyr) & $t_{\rm psr}$ & 10 & Assumed\\
Breaking index & $n$ & 3 & Assumed \\
Initial spin down timescale (kyr) & $\tau_0$ & 2 & Assumed \\
\\
Initial spin down luminosity (${\rm erg\ s^{-1}}$) & $L_0$ & $1.5\times10^{38}$ 
& This work \\
Break energy & $\gamb$ & $2.5\times10^5$ & This work \\
Low energy index & $\alpha_1$ & 1.5 & Fixed \\
High energy index & $\alpha_2$ & 2.55 & This work \\
Shock radius fraction & $\varepsilon$ & 0.1 & This work \\
Magnetic fraction & $\eta$ & 0.09 & This work \\
\\
CMB temperature (K) & $T_{\rm CMB}$ & 2.73 & Fixed \\
CMB energy density (${\rm eV\ cm^{-3}}$) & $w_{\rm CMB}$ & 0.26 & Fixed \\
IR temperature (K) & $T_{\rm IR}$ & 40 & Fixed \\
IR energy density (${\rm eV\ cm^{-3}}$) & $w_{\rm IR}$ & 1 & Fixed \\
SL temperature (K) & $T_{\rm SL}$ & 4000 & Fixed \\
SL energy density (${\rm eV\ cm^{-3}}$) & $w_{\rm SL}$ & 2 & Fixed \\
\enddata
\end{deluxetable*}

\begin{acknowledgements}
	We thank the referee for very detailed comments, which 
	greatly helped improve this paper.
This research is supported by the Basic Research Program of Yunnan Province
No. 202201AS070005 and the National Natural Science Foundation of China 
(12273033). D. Z. thanks the support of the Basic Research Program 
of the Education Division of Yunnan Province No. 2022Y055, and Z.W. 
	acknowledges the support by the Original Innovation 
Program of the Chinese Academy of Sciences (E085021002). X.Z. and Y.C. thank the
	support of National Key R\&D Program of China under nos. 2018YFA0404204 and 2017YFA0402600 and NSFC grants under nos. U1931204, 
12173018, 12121003, 11773014, and 12103049.

\end{acknowledgements}

\bibliography{gpwn}
\bibliographystyle{aasjournal}



\end{document}